# Real-time *In Situ* Electron Spin Resonance Measurements on Fungal Spores of *Penicillium digitatum* during Exposure of Oxygen Plasmas

**Kenji Ishikawa[1,3], Hiroko Mizuno[1], Hiromasa Tanaka[1], Kazuhiro Tamiya[2], Hiroshi Hashizume[2], Takayuki Ohta[2], Masafumi Ito[2], Sachiko Iseki[1], Keigo Takeda[1], Hiroki Kondo[1], Makoto Sekine[1], and Masaru Hori[1]**

[1] Nagoya University, Furo-cho, Chikusa, Nagoya, Aichi 464-8603, Japan

[2] Meijo University, 1-501 Shiogamaguchi, Tenpaku, Nagoya, Aichi 468-8502, Japan

We report the kinetic analysis of free radicals on fungal spores of *Penicillium digitatum* interacted with atomic oxygen generated plasma electric discharge using real time *in situ* electron spin resonance (ESR) measurements. We have obtained information that the ESR signal from the spores was observed and preliminarily assignable to semiquinone radical with a g-value of around 2.004 and a line width of approximately 5G. The decay of the signal is possibly linked to the inactivation of the fungal spore. The real-time *in situ* ESR has proven to be a useful method to elucidate plasma-induced surface reactions on biological specimens.

[3] Author to whom correspondence should be addressed. Electronic mail: ishikawa.kenji@nagoya-u.jp

Electrical discharges of gases produce state called 'low temperature plasma' including electrons, ions and radicals and having only high temperature for electrons but for background gases. Recently, the potential applications of non-equilibrium atmospheric pressure discharge in biology and medicine has grown significantly.[1] Plasma interaction with living tissues and cells has not been understood to the extent of comprehensive biological and physical mechanisms, due to the complexity of both the plasma and tissue. Therapeutic effects, such where plasmas can trigger a complex sequence of biological responses in tissues and cells, are indeed complex. In addition, mammalian cells, but not bacteria, repair easily with participation of adducted-deoxyribonucleic-acid (DNA) intracellular activation.[2] Many interpretations of the biological responses for disinfection by plasma discharges have been reported, which has indicated that death is accompanied by (1) lethal oxidation of membranes, (2) electrostatic disruption of cell membranes, such as cell lysis and fragmentation, and (3) biological responses such as apoptosis.[2] A biological organism is catalytically oxidized by positive and negative ions both inside and outside of the biological organism in the presence of oxygen and reactive oxygen species. Therefore, not only ultraviolet photons, but also chemical species including (1) positive and negative ions, (2) reactive oxygen species, (3) hydrogen-peroxide, and (4) ozone are considered to be provide function of modification on biological organism.[3-5] Kong *et al.* identified oxygen atoms as a major contributor in plasma inactivation,[6] and Nagatsu *et al.* reported that spore forming bacteria, *Geobacillus stearothermophilus*, are inactivated by surface wave excited oxygen plasmas.[7]

*Penicillium digitatum* is a post-harvest disease, which is difficult to inactivate, causes the green mold of citrus fruits, and differs slightly from other microorganisms such as *Escherichia coli* and *Bacillus subtilis* that are typically found on medical instruments. Cerioni *et al.* have reported cellar damages of oxidative treatments by observing transmission electron micrographs of conidia, that is, spore of *Penicillium digitatum*.[8] In our recent publication, fungal spores of *Penicillium* *digitatum* were disinfected by applying atmospheric pressure plasmas of argon gas containing trace oxygen. The results of this experiment indicated there was no contribution of ultraviolet radiation and no inactivation effects of nitric oxide (NO) or ozone ($O_3$).[9] The rates of inactivation were shown a good correlation with density of atomic oxygen regarded as a major contributor.[10]

The free radical has been historically studied due to the identification of biological metabolic mechanisms.[11] In a number of studies, the free radical content of tissues was found using the electron spin resonance (ESR) method. For over 40 years, there have been many studies for the ESR signals of various spores and other biological tissues.[12] In particular, the reduced-form molecule such as semiquinone (QH*; the * means radical state) plays a role to protect from oxidation by incoming reactive oxygen species. Evidently, the protective mechanism for the direct reaction of the oxidized-form molecule must be different from the radical scavenger (antioxidant) mechanism proposed for indirect action. However there still remains a need to clarify interaction of biological specimens with low-temperature-plasmas itself not but radiation damages.

Recently, the real-time *in situ* ESR method has been progressively developed to observe the chemical reactions of materials during irradiation of plasmas.[13-15] We emphasize that not only surface modification of poly(tetrafluoroethylene) (PTFE) film with hydrogen discharged plasma[15], but also biological applications of plasma can be applied. Namely we attempt to reveal interactions with a focus on a boundary layer between the low-temperature plasma and the material (biological) surfaces, and to comprehensively understand a mechanism of interactions on microorganisms to distinguish both individual and synergistic roles for ions, radicals, photons, etc. generated by plasmas. We have used this technique to observe free radicals on *Penicillium digitatum*.

In this study, we have attempted to elucidate the mechanism on the plasma interaction of living tissues with reactive oxygen species. Thus, the interaction of *Penicillium digitatum* with plasmas has been a substantial questioned. We report on the real-time *in situ* observation of the interaction of a bacterial cell with electrically discharged plasma.

Spores of *Penicillium digitatum* of 10 mg were immersed and dispersed into 1 ml of sterilized water added 1% of a surface active agent (Tween 20; Polyoxyethylene (20) sorbitan monolaurate, CAS No. 9005-64-5) to avoid clumping of the spores. Samples were prepared on synthetic quartz glass plates (3×7 mm$^2$ size and 0.5 mm thickness). The aqueous dispersed liquid of 1 to 4 µl was dropped on the quartz glass plate, which resulted in the deposition of spores on the quartz glass surface, followed by drying.

ESR measurements were conducted using a standard X-band (9 GHz) spectrometer (Bruker Biospin, EMX plus) with a microwave resonator (Bruker Biospin, ER4119HS-W1). ESR spectra were recorded using a microwave power of 0.2 mW (note that signals were saturated at above 2 mW.), a field modulation amplitude of 0.05 mT, and a modulation frequency of 100 kHz. All experiments were conducted at room temperature. For the real time measurements, the field modulation amplitude was changed to higher values of 0.5 mT for improving signal-to-noise ratio. The quartz glass plate with deposited spore sample was inserted into a synthetic quartz glass tube with a diameter of approximately 9 mm, which was set inside of the ESR cavity in the down flow region, typically 20 cm from the plasma discharge.

A microwave (2.45 GHz) power supply of 50 W from a generator (Horonix MR-202) was used to generate plasma inside of the quartz tube. Typically oxygen gas was flowed into the quartz tube at a flow rate of 15 sccm, and the pressure was maintained at approximately 15 Pa in the downflow region. It should be noted that only oxygen plasmas were used in this study.

Individual contributions from atomic oxygen and light of plasma emissions have been studied in a similar manner using pallets for plasma evaluation (PAPE).[15-17] The distance from the discharge was sufficiently large, so that charged particles such as electrons and ions were immediately diminished by recombination. We confirmed experimentally no electron cyclotron resonance originating from surviving electrons as far as position of the ESR cavity, i.e. it can be ignorable to affect the electron related phenomena including Penning ionization from highly excited long-lived metastables. Therefore, there was no ion bombardment effect observed at the down-flow region. To limit the interactions of plasma emissions with the sample, a U-shaped quartz tube and light shade were used, as shown in Figure 1. Neutral species such as atoms and radicals are relatively long-lived; therefore, a sufficient amount of neutral species were transported for interaction with the spores. In fact, the exposure of atomic oxygen was evidenced by detection of the ground state of atomic oxygen ($^3P_j$; j=0,1,2) by the ESR. Unless otherwise stated, all experiments were conducted to irradiate samples with neutrals and radicals.

Working cultures were prepared by collecting and suspending the spores with 1 ml of sterilized water added 1% of Tween 20 and harvested uniformly on 4% potato dextrose ager (PDA) broth in petri dishes with a diameter of 90 mm. Spore cultures were kept on the PDA broth and developed for a period of three days at 25° C. The decimal reduction time, D-value, which is the exposure time required under specified set of conditions to cause one logarithmic reduction (or 90%) of the initial population was characterized. The initial concentration of spores (dilution of $10^5$) was approximately $10^8$ to $10^9$ colony forming units (cfu)/ml. Also the cultures were conducted for samples prepared for different periods of every two minutes of the treatments, which were exposed to atomic oxygen generated plasmas. Assay for D-value was characterized with taking care of no effect of blowout of spores during flowing of gases for the treatments. The experiments avoided the blowout with a use of poly(tetrafluoroethylene)(PTFE) membranes placed on front and back of the glass plate deposited the spores.

Figure 2 shows the experimental ESR spectra. ESR signals from fungal spores of *Penicillium digitatum* are shown in Figure 2c. The spectra are represented by Gaussian signals located at g-value of 2.0040 with a line width of approximately 5 G. No signal was detected from the Tween 20 as shown in Figure 2a. Figure 3 shows the microscopic images of the fungal spores and sprouts. Following the germination of fungal filaments (colorless) from the spore (dark greenish color) after several days, no ESR signal due to germination was detected (Figure 2b). During the flow of oxygen gas, ESR signals due to gaseous oxygen molecules ($^3\Sigma_g^-$) [18] were observed in the ESR spectra, as indicated in Figure 2d. To identify the ESR signal detected here, we firstly made a preliminarily speculation that the intracellular semiquinone radical (QH*) is a candidate for the ESR signal source.[19-23] Powder of 1,4-Benzoquinone (CAS No. 106-51-4), which contains complex-form hydoroquinone-quinone was measured and Figure 2e shows ESR spectrum is represented by overlapping Gaussian signals located at 2.0040, 2.0072, 2.0042 and 2.0013 and line widths of 5.0, 5.5, 3.0 and 6.0 G, respectively.(Fig. 2f) This indicates that the QH* were assignable to the case of *Penicillium digitatum*, and that the detected ESR signals were due to the fungal spores of *Penicillium digitatum*.

The real-time signal decay was measured during irradiation with atomic oxygen. Figure 4a shows the ESR spectra taken at each irradiation time. Figure 4b shows the temporal change of the integrated ESR intensity with respect to the exposure time. When volumes of the dispersed liquid were changed 1,2 and 4 μl, the ESR intensities were in proportion to the number of spores. After irradiation from the start of plasma discharge, the ESR signals due to *Penicillium digitatum* were clearly decreased. Since the U-shaped tube and light shade, it was concluded that this decay resulted from the individual contribution of atomic oxygen rather than from plasma emissions.[15] The spore appearance was notably changed to almost colorless mycelia from dark-greenish spores and the synchronous change in decoloration was accompanied by detection of an ESR signal. (The fungal spores are not, of course, inactivated to be kept inside vacuum chamber. There is no disruption of spores, thus it was evident that no substantial etching of the spore walls occurred in this study.) Notably rates for the ESR intensities may represent as similar to the dose-response (logistic) curve, and the intensities were saturated a constant value for longer periods.

At present, we speculated that membranes function to protect from oxidative stress, so that a thick spore wall made of polysaccharide could play an important role in the radical scavenging behavior. There is one idea that the superoxide anion radical ($O_2^-$*) can not penetrate lipid membranes, but the superoxide can penetrate membranes.[11] The ground state of the fatty acid main components of membranes has singlet multiplicity and reaction with oxygen is spin forbidden, because the ground state of oxygen has triplet multiplicity. Molecular oxygen as a triplet species cannot react at a significant rate with most biomolecules.[11] Similarly, oxygen atom singlet species may react with the spore wall; therefore, empirical evaluation suggests that the peroxidation reaction of the membrane and lipid involves a mechanism that circumvents the spin barrier.[24] One possibility is reaction via free radical mechanisms involved with chain reactions. The radical initiated autoxidation of lipids occurs *in vivo*; firstly lipid (LH) initiated L* and propagation reaction, L* + $O_2$ → LOO*, then LOO* + LH → LOOH + L*, and finally termination as follows, LOO* + L* → LOOL, and LOO* + LOO* → LOOL + $O_2$.[25] In addition to the radical scavenging processes by antioxidants on polysaccharide or phenolic compounds, lipid peroxidation (OO*) in biological systems is also a good example,[26] and on lipid and membrane, the oxidation protection mechanism is valid. Hence we speculate that an excess of oxidants such as $O_2^-$* are supplied inside cell, so that then reductive-state molecules react to form products. Considering the general knowledge of energy metabolism, QH* as a reductase is oxidized to form hydroquinone ($QH_2$) by $O_2^-$*. This disproportionation reaction of QH* with Q and $QH_2$ products continually destroys and reforms the free radicals. Following this assumption, the rate-limiting step being peroxidation of the membrane to generate oxidants, *i.e.*, the disproportionation reaction. The oxidant could then react with an abundance of intracellular stable radical scavengers, such as the QH* reductase. Gaseous reactive oxidants reach at the surface, then diffuse into the spore wall, and react cooperatively the incoming oxidants with the chemicals such as semiquinone, then the signal decrease may behave to be logistic equation,[27] the ESR intensity, $I$, can be simply formulated by

$$I = I_0 + \frac{I_s - I_0}{1 + (k/t)^n},$$

where $I_0$ is the initial value, $I_s$ is the saturated value, $k$ is a coefficient for time-behavior, and $n$ is a cooperative factor. Here the fitting represents well with agreement of experimental values and we typically obtained values for the $n$ is 3 and $k$ is ranged in 10 to 12 min. Thus we propose that QH* in a biologically vital metabolism is eventually oxidized and terminated indirectly by irradiation with atomic oxygen. At present the interpretation however requires further clarification.

Moreover, Figure 5 shows the number of surviving spores as a function of plasma treatment time. D value of the spores was approximately 5 min. The number of spores surviving the irradiation was clearly decreased and this results was in agreement with other report.[10] However we observed that the rate was saturated at two or three order of decrease after approximately 10 min. Compared with the ESR results as shown in Fig. 4, even though differences in the inactivation behavior and the ESR intensities, we believe that the ESR signals are correlated with the inactivation. At least, the real-time *in situ* ESR measurement system is an effective approach to determine reactions related with the radical.

In summary, the ESR signal from *Penicillium digitatum* spores was observed at a g-value of *ca*. 2.004 with a line width of *ca*. 5G. The signal was preliminarily assigned to a stable free radical, such as intracellular semiquinone. After germination of the spore, the signal was diminished with the change in the shape and color of the mycelia. It is noteworthy that the ESR signal was decreased when only oxygen plasma was discharged. The real-time *in situ* ESR measurement signal decayed during O atom irradiation. We have obtained information regarding the reaction mechanism with free radicals and this real-time *in situ* ESR method has proven to be a useful method to elucidate plasma-induced surface reactions on biological systems.

**Acknowledgement**

This work was supported in part by the Knowledge Cluster Initiative (Second Stage) of the Tokai Region Nanotechnology Manufacturing Cluster. The authors would like to thank Naoya Sumi, for technical support.

**FIGURE CAPTIONS**

Figure 1. Schematic diagram of the experimental setup for real-time *in situ* ESR measurements. Only atomic oxygen can be irradiated.

Figure 2. ESR spectra from (a) no signal from Tween 20 only, (b) sprouts of *Penicillin digitatum*, (c) fungal spores of *Penicillin digitatum* before plasma exposure, (d) during $O_2$ plasma exposure, (e) powder of 1,4-Benzoquinone and (f) deconvolved spectrum of (e).

Figure 3. Microscopic images for (a) fungal spores and (b) sprouts of *Penicillin digitatum*.

Figure 4. Temporal changes during exposure to atomic oxygen in $O_2$ plasma; (a) a series of ESR spectra and (b) ESR intensity of *Penicillin digitatum* as a function of atomic oxygen exposure time. Measurements were conducted *in situ* at real time.

Figure 5. Number of survivor as a function of treatment time for exposure to atomic oxygen.

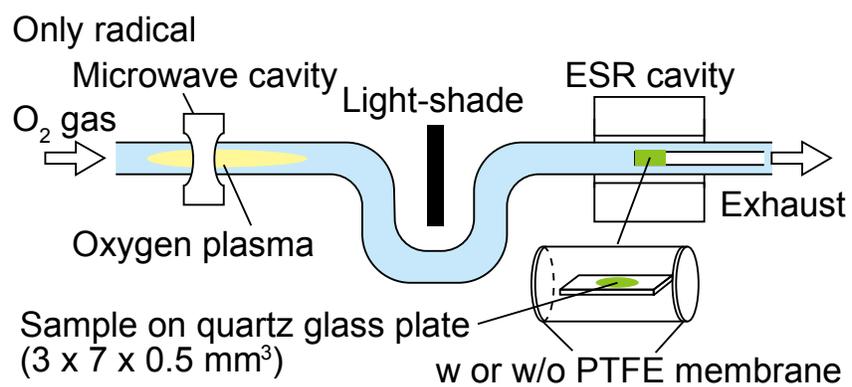

Figure 1 Ishikawa

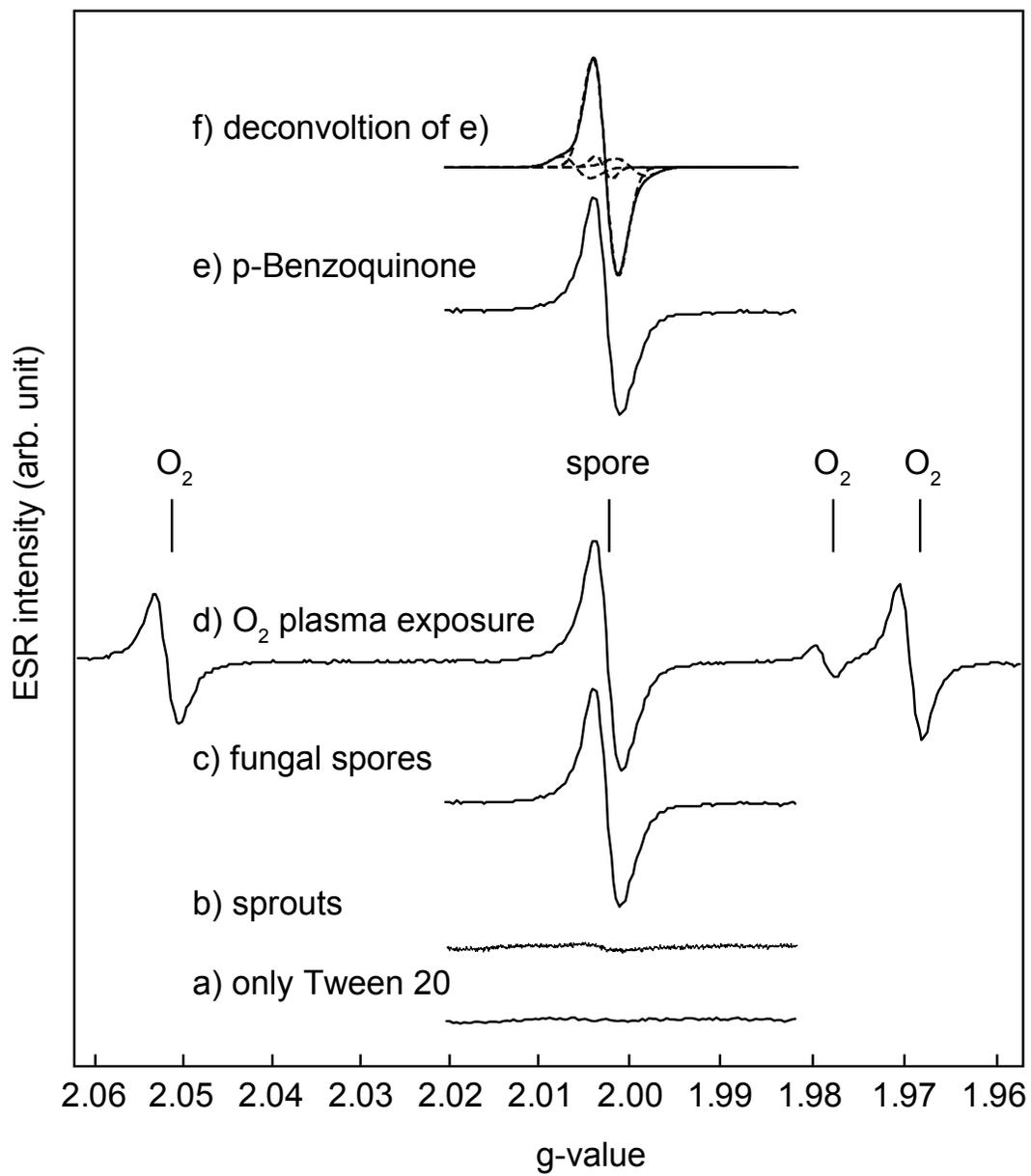

Figure 2 Ishikawa

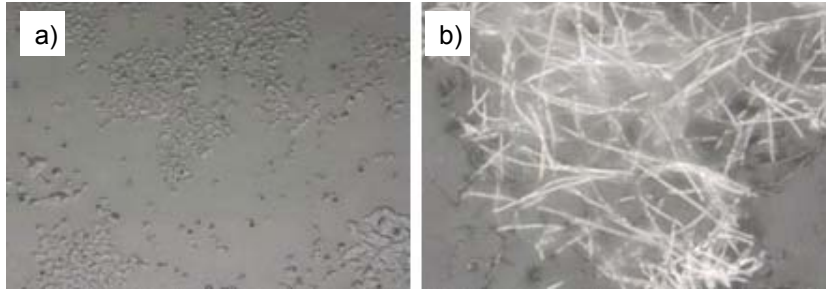

Figure 3 Ishikawa

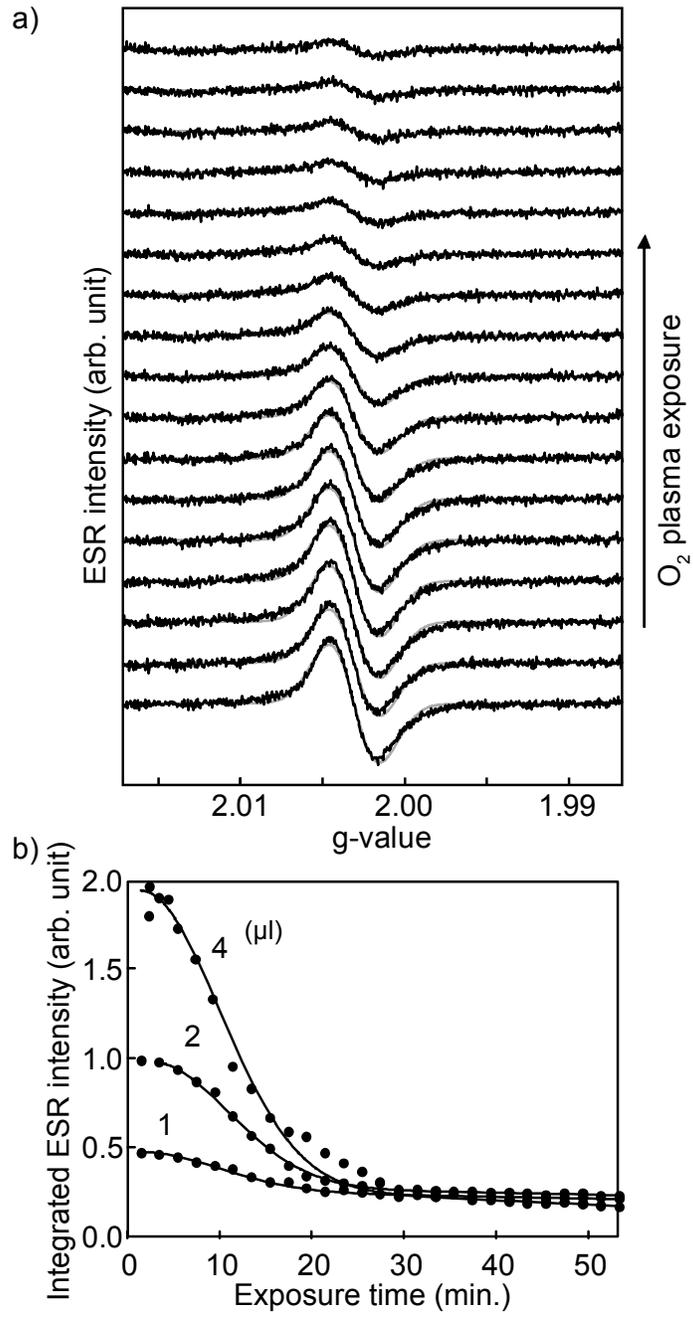

Figure 4 Ishikawa

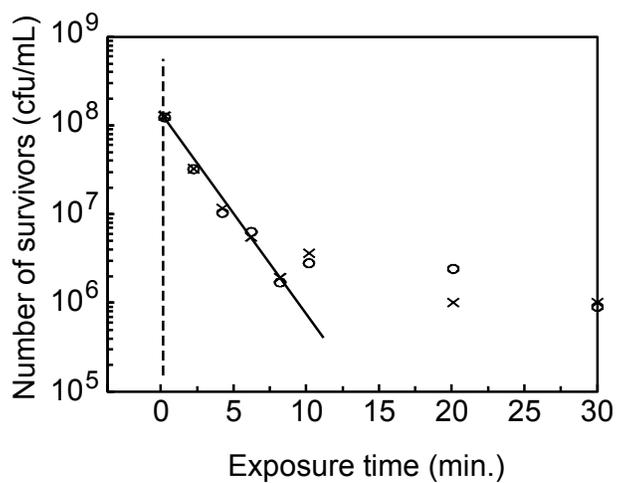

Figure 5 Ishikawa